\documentclass[12pt]{article}
\pdfoutput=1
\usepackage{euscript,amsmath,amsthm,amsfonts,amssymb}
\usepackage{mathtools}
\usepackage{subfig}
\usepackage{float}
\usepackage[margin=1in]{geometry}
\usepackage{graphicx}
\usepackage{outlines}
\usepackage[dvipsnames]{xcolor}
\usepackage{jheppub}
\usepackage{cleveref}
\newcommand{\ket}[1]{|{#1}\rangle}

\newcommand{\bra}[1]{\langle{#1}|}

\title{\boldmath $SU(N)_{1}$ Chern-Simons theory, the Clifford group, and Entropy Cone}
\author{Howard J. Schnitzer}
\affiliation{Martin Fisher School of Physics, Brandeis University, Waltham, Massachusetts 02453, USA}
\preprint{BRX-TH-6662}
\emailAdd{schnitzr@brandeis.edu}
\abstract{Entropy cones for $SU(N)_{1}$ Chern-Simons theory are discussed. It is shown that stabilizer states can be constructed from topological operators in $SU(N)_{1}$ for $N$ odd prime, but not for $SU(N)_{K}$; $K \geq 2$. This implies that the topological entropy cone is properly contained in the stabilizer entropy cone for $SU(N)_{K}$; $K \geq 2$.

}
\begin{document}
\maketitle
\flushbottom

\section{Introduction}
There is a great deal of interest in the entropy cones of various theories; in particular holographic entropy cones \cite{Swingle:2012wq,Bao:2015bfa,Bao:2015boa,Freedman:2016zud,Hubeny:2018trv,Hubeny:2018ijt,Cuenca:2019uzx,He:2019ttu,Akers:2019gcv,He:2020xuo}. By contrast the entropy cones for topological theories, such as Chern-Simons theories have received less attention.

In general, quantum systems can be describe by N-partite subsystems, together with entanglement entropies of all marginal reduced density operators. This defines a vector in $\mathbb{R}_{2N}$, where the closure of all such vectors is a convex cone. A quantum entropy cone is described by the universal inequalities satisfied by the von Neumann entropies of a multi-partite system. There are various entropy cones of interest in generic quantum theories, as well as those relevant to Chern-Simons theories. In particular, we focus on
\begin{outline}[enumerate]
\1 stabilizer entropy cones obtained from qudit states, and the generalized Pauli group used to construct stabilizer states \cite{Linden:2013kal}
\1 topological entropy cones, which involve those stabilizer states which can be constructed from topological operators of the theory \cite{Salton:2016qpp}.
\end{outline}
For example, in Section 3 we show that stabilizer states can be constructed topologically for qudits in Chern-Simons $SU(N)_{1}$ theory ($N$ odd prime), but not for $SU(N)_{K}$; $K \geq 2$. The issue is that in the latter case, the Heisenberg group is not topological, so that the topological cone is contained in the stabilizer cone.

Although holographic entropy cones relevant to holographic theories are outside the scope of this paper, it has been conjectured that the holographic entropy cone can be identified with the topological entropy cone in holographic theories based on tensor networks and bit string models \cite{Cui:2018dyq} for the RT relation. This conjecture has been questioned by Akers and Rath \cite{Akers:2019gcv}, who argue that a holographic CFT requires a large amount of tripartite entanglement, as GHZ$_{3}$ tripartite entanglement is not sufficient, but that tripartite W$_{3}$ states could be considered as well since tensor networks have flat Renyi entropies \cite{Cui:2018dyq,Hayden:2016cfa}. These issues can be considered for Chern-Simons theories modeled by tensor networks \cite{Nezami:2016zni}. Since GHZ$_{3}$ is a stabilizer state for qubits, but W$_{3}$ is not, the consideration of Chern-Simons theories can be instructive.

In section \ref{sec:2} we review some properties of quantum entropy relations, and entropy cones. In section \ref{sec:3} we discuss the Clifford group for $SU(N)_{1}$ [$N$ odd prime for simplicity], and show that stabilizer states can be constructed topologically, but not for $SU(N)_{2}$, and by implication not for $SU(N)_{K}$; $K \geq 2$. For example, both $SU(3)_{1}$ and $SU(2)_{2}$ have qutrit basis, but $SU(2)_{2}$ has stabilizer states which are not topological.

In section \ref{sec:4} we consider Chern-Simons $SU(N)_{1}$ theories as described by large $N$ tensor methods \cite{Nezami:2016zni}, which results in a flat entanglement spectrum dominated by Bell pairs and $\mathcal{O}(1)$ GHZ$_{3}$ tripartite states. Section \ref{sec:5} has concluding remarks, and suggestions of issues for further consideration.

\section{Entropy relations and quantum entropy cones}\label{sec:2}
\subsection{Entropy relations}
In this subsection we summarize some relevant entropies \cite{Bao:2015bfa,Bao:2015boa,Freedman:2016zud,Hubeny:2018trv,Hubeny:2018ijt,Cuenca:2019uzx,He:2019ttu}. Given any two subsystems $A$ and $B$ of a theory, with $A\cup B \equiv AB$, the subadditivity for von Neumann entropies
\begin{equation}
S(A) + S(B) \geq S(AB)
\end{equation}
holds for any quantum system and is equivalent to the positivity of mutual information, i.e.
\begin{equation}
I_{2}(A:B) = S(A) + S(B)-S(AB) \geq 0
\end{equation}
For a division into three subsystems, $A$,$B$ and $C$, there is strong subadditivity (SSA)
\begin{equation}
S(AB) +S(BC) \geq S(B) +S(ABC)
\end{equation}
which is equivalent to the monotonicity of mutual information
\begin{equation}
I_{2}(A:BC) \geq I_{2}(A:B)
\end{equation}
and is universal.

However there are relations which are not universal, such as the monogamy of mutual information (MMI)
\begin{equation}
S(AB)+S(BC)++S(CA) \ge S(A) +(S(B)+S(C)+S(ABC)
\end{equation}
which can be expressed as 
\begin{equation}
I_{3}(A:B:C)=S(A)+S(B)+S(C)-S(AB)-S(BC)-S(CA)+S(ABC) \leq 0
\end{equation}
An example of non-universality is the 4-partite GHZ$_{4}$ state which is a stabilizer state that violates MMI.

The Ingleton inequality \cite{Linden:2013kal} [ING] is not obeyed by a general quantum states. It is 
\begin{equation}
ING(AB:CD) = I_{2}(A:B|C)+I_{2}(A:B|D)+I_{2}(C:D)-I_{2}(A:B) \geq 0
\end{equation}
with the definition
\begin{equation}
I_{2}(A:B|C)=S(AC)+S(BC)-S(ABC)
\end{equation}
where $A$,$B$,$C$ and $D$ are generically pairwise disjoint subsets. If MMI holds then ING is satisfied \cite{Bao:2015bfa,Linden:2013kal}. A 4-partite quantum cone can be stabilizer represented iff ING is satisfied \cite{Linden:2013kal}. For a stabilizer tensor network, 4-partite entanglement is 1-1 between the building blocks of the 4-partite stabilizer cone. Any \emph{pure} stabilizer state in a 5-partite system implies a 4-partite reduced state which holds for ING \cite{Linden:2013kal}. More generally, any pure stabilizer state for a $(N+1)$ partite system generates a N-partite entropy vector which satisfies the Kinser family of inequalities $K[N]$ which reduce to ING for $N=4$. That is $K[4] =$ ING \cite{Linden:2013kal}.

There are a number of distinguished pure states discussed by Bengtsson and Zyczkewski \cite{bengtsson2016brief}. For 3 qubits GHZ$_{3}$ is a stabilizer state. Another category of multipartite entanglement states are the absolutely maximally entangled states $\equiv$ AME, which exist for $K=2,3,5$ and 6 qubits, but not for $K=4$ or $K\geq6$ qubits. For $K=3$ qubits the AME state is GHZ$_{3}$, while for $K=5$ qubits, there is the stabilizer state
\begin{equation}
\Omega_{5,2} = \frac{1}{2\sqrt{2}}\{\ket{00000} +\ket{00011} +\ket{01100}-\ket{01111}+\ket{11010}+\ket{11001}+\ket{10110}-\ket{10101}\}
\end{equation}

There is an AME state for 4-qutrits in $SU(3)$,
\begin{equation}
\ket{\Phi^4_3} = \ket{0000} + \ket{0112}+\ket{0221}+\ket{1011}+\ket{1120}+\ket{1202}+\ket{2022}+\ket{2101}+\ket{2210}
\end{equation}
which is a stabilizer state. In general there are a large number of stabilizer states which exhibit multi-particle entanglement. An open question is which of the various stabilizer states can be constructed topologically in Chern-Simons theory? This issue is considered in section 3 for $SU(N)_{1}$; $N$ odd prime and $SU(N)_{K}$; $K\geq2$. A necessary condition is that the Heisenberg group be constructed topologically. We show explicitly in section \ref{sec:3} that the Clifford group can be constructed from topological operators for $SU(N)_{1}$; $N$ odd prime. For $N$ even, the work of Farinholt \cite{Farinholt_2014} is likely to be useful.

\subsection{Quantum Entropy cones: $c^Q_n$}
A number of entropy cones have been discussed recently. These are
\begin{equation*}
\begin{split}
c^Q_n &= \text{quantum entropy cone,}\\
c^h_n &= \text{holographic cone,}\\
c^S_n &= \text{stabilizer cone,}\\
c^t_n &= \text{topological cone,}\\
c^H_n &= \text{Hypergraph cone.}
\end{split}
\end{equation*}

In section \ref{sec:3} we show that $c^t_n \subseteq c^S_n$ for Chern-Simons $SU(N)_{K}$ theory. In general $c^S_{n}\subseteq c_{n}^{Q}$. Not all stabilizer states can be constructed from topological operations. However $c^t_{n}= c_{n}^{S}$ for $SU(N)_{1}$; $N$ odd prime, while $c^t_{n}\subset c_{n}^{S}$ for $SU(N)_{K}$; $K\geq2$.

$c^H_{4}\subseteq c_{4}^{S}$ \cite{Bao:2020zgx}, while $c^H_{n}\subset c_{n}^{S}$; $n\geq5$ \cite{Walter:2020zvt,Bao:2020mqq}. The holographic entropy cone is equivalent to the graph entropy cone for any number of parties \cite{Bao:2015bfa}.

It is a developing issue to understand the role of the various entropy cones in topological theories, and in Chern-Simons theory in particular. This is the main focus of secs. \ref{sec:3} and \ref{sec:4} of this paper.

\section{Stabilizer cone for $SU(N)_{1}$}\label{sec:3}
This discussion is a continuation of section 4 of Salton, et. al. \cite{Salton:2016qpp}, where we give a parallel construction of the generalized Pauli group, which can be prepared by topological operations for $SU(N)_{1}$, $N$ odd prime, as a higher dimensional generalization of the Pauli group, in analogy with the general discussion of Gottesman \cite{1999} for qudits, with $d$ an odd prime.

Consider $SU(N)_{1}$, with $N$ an odd prime. Representations can be described by a single column Young tableau, with zero, one, two, ... , $(N-1)$ boxes. The modular transformation matrix $S_{ab}$ can be viewed as a basis change
\begin{equation}
\ket{a} = \sum^{N-1}_{b=0}S_{ab}\ket{b},\quad a=0\text{ to }N-1,
\end{equation}
where $S_{ab}$ can be prepared topologically, in analogy with Salton, et. al. \cite{Salton:2016qpp}. The fusion tensor satisfies
\begin{equation}\label{eq:fusion0}
N_{ab}^c:\quad a+b=c\mod N
\end{equation}
so that
\begin{equation}\label{eq:fusion}
N_{a,1}^b=\delta_{a+1,b}\mod N,
\end{equation}
where $N_{ab}^{c}$ can also be constructed topologically as in Salton, et. al. \cite{Salton:2016qpp}. The modular transformation matrix can be constructed from equation (2.6) of Mlawer et. al. \cite{Mlawer:1990uv} using
\begin{equation}
S_{\sigma(a),b}=e^{\frac{-2\pi i r(b)}{N}}S_{ab}
\end{equation}
where $r(b)$ is the number of boxes in the representation $b$, while $\sigma(a)$ is obtained from $a$ by adding a single vertical box to the representation. Further $S_{ab}=S_{ba}$ from equations (3.3) and (3.4) of Mlawer et.al. \cite{Mlawer:1990uv}. It is straightforward to show that for
\begin{equation}
SU(2)_{1}: \quad S_{ab}=\frac{1}{\sqrt{2}}
\begin{pmatrix}
1&1\\
1&-1
\end{pmatrix}
\end{equation}
\begin{equation}
SU(3)_{1}: \quad S_{ab}=\frac{1}{\sqrt{3}}
\begin{pmatrix}
1&1&1\\
1&w^{-1}&w^{-2}\\
1&w^{-2}&w^{-4}
\end{pmatrix}
\end{equation}
where $w=e^{\frac{2\pi i}{3}}$ is a primitive third root of unity. The general result for $SU(N)_{1}$ is
\begin{subequations}\label{eq:transformmatrix}
\begin{equation}
SU(N)_{1}: \quad S=\frac{1}{\sqrt{N}}
\begin{pmatrix}
1&1&1&\cdots&1\\
1&w^{-1}&w^{-2}&\cdots&w^{-(N-1)}\\
1&w^{-2}&w^{-4}&\cdots&w^{-2(N-1)}\\
1&w^{-3}&w^{-6}&\cdots&w^{-3(N-1)}\\
\vdots&\vdots&\vdots&\ddots&\vdots\\
1&w^{-(N-1)}&\cdots&\cdots&w^{-(N-1)^2}
\end{pmatrix}
\end{equation}
where $w^{N}$=1. This can be written as
\begin{equation}\label{eq:7b}
S = \frac{1}{\sqrt{N}}\sum_{a=0}^{N-1}\sum_{b=0}^{N-1}w^{-ab}\ket{a}\bra{b},
\end{equation}
\end{subequations}
so that for $SU(N)_{1}$ the modular transformation matrix coincides with the qudit Hadamard matrix. Note that in \eqref{eq:transformmatrix}, each row sums to zero, except for row zero.

From this data one can construct the generalized Pauli group for $SU(N)_{1}$ \cite{Schnitzer:2019icr}, where for simplicity it is useful to restrict $N$ to an odd prime, following Gottesman \cite{1999}. For even $N$, see Farinholt \cite{Farinholt_2014}. The generalized Pauli group for qudits is generated by tensor products of $X$ and $Z$, defined by
\begin{equation}\label{eq:3.8}
X\ket{a} = \ket{a+1,\mod N}
\end{equation}
\begin{equation}\label{eq:3.9}
Z\ket{a} = w^{a}\ket{a}
\end{equation}
and
\begin{equation}\label{eq:3.10}
XZ=w^{-1}ZX
\end{equation}
where $w$ is the $N$-th primitive root of unity [In our case, $w^{N}=1$; $N$ an odd prime] Identify $X$ with the fusion tensor \eqref{eq:fusion}. In detail
\begin{equation}\label{eq:3.11}
Z_{ad} = \sum_{b,c=0}^{N-1}S_{ab}N_{b,1}^c(S^{\dagger})_{cd},
\end{equation}
so that
\begin{equation}\label{eq:3.12}
Z_{ac} = \sum_{b=0}^{N-1}[S_{ab}(S^{*})_{b+1,a}]\delta_{ac}
\end{equation}
which one can verify by explicit calculation. Therefore, the generalized Pauli group for $SU(N)_{1}$, $N$ odd prime can be constructed topologically by path integration in analogy with equation (8) of Salton, et. al. \cite{Salton:2016qpp}.

$SU(N)_{1}$ Chern-Simons theory is the level-rank dual of $U(1)_{N}$ Chern-Simons theory. As a result, the S-matrix \eqref{eq:7b} is identical to that of equation (16) of \cite{Salton:2016qpp}, and the fusion matrix \eqref{eq:fusion0} coincides with equation (7) of \cite{Salton:2016qpp}. Therefore, the single qudit Pauli group generated is identical, as can be verified by comparing \cref{eq:3.8,eq:3.9,eq:3.10,eq:3.11,eq:3.12} with \cite{Salton:2016qpp}. As a consequence, $X$ and $Z$ are both level-rank invariant, and the stabilizer states of both theories are identical. The level-rank dual of Theorem 1 of \cite{Salton:2016qpp} implies that for $SU(N)_{1}$, $N$ an odd prime, one can prepare any stabilizer state in the n-torus Hilbert space. Further, for $SU(N)_{1}$ $N=5$ e.g., one conjectures that one can only prepare stabilizer states.\footnote{The level-rank dual of link variables discussed in \cite{Mlawer:1990uv}, is needed for the level-rank dual of Theorem 1 of \cite{Salton:2016qpp}.}

The Clifford group is the set of operators which leave the generalized Pauli group $P$ invariant under conjugation. That is, it is the normalizer $N(P)$ of $P$ in the unitary group $U(N^n)$ for $n$-qudits. The stabilizer states are the eigenstates of the maximal invariant subgroup of the Pauli operators. This implies that the stabilizer entropy cone, in general is a subspace of the quantum entropy cone. A central issue is whether these cones can be constructed by topological operations. As we have shown for $SU(N)_1$, $N$ odd prime, stabilizer states can indeed be constructed topologically.

For $SU(N)_K$; $K\geq2$, the Pauli group cannot be constructed topologically as above. For example, for $SU(2)_{2}$ qutrits, the fusion tensor is
\begin{equation}
N_{ab}^c:\quad a+b=c\mod 2,\text{ where }a,b=0,1\text{ or }2.
\end{equation}
In this case the analogues of the above  \cref{eq:3.8,eq:3.9,eq:3.10,eq:3.11,eq:3.12} are not applicable. Therefore, for $SU(N)_{K}$; $K\geq2$ the stabilizer entropy cone is larger than the topological entropy cone.

\section{Large $N$ tensor networks for $SU(N)_{1}$}\label{sec:4}
Consider Chern-Simons theory on a three-dimensional manifold $M$, whose boundary $\partial M$ is that of a n-torus Hilbert space. A Riemann surface without boundary will be identified with $\partial M$. Generically a genus $g$ Riemann surface without boundaries can be represented by a polygon with $4g$ identified sides. For example, the genus 2 surface can be represented by an octagon with identified sides
\begin{equation}
aba^{-1}b^{-1}cdc^{-1}d^{-1}
\end{equation}
However, for large $N$ $SU(N)_{1}$ Chern-Simons theories constructed from a tensor network, the Renyi entropies are flat, so that one need not specify the genus of the polygon representation of $\partial M$. [This section is parallel to that of Salton, et. al. \cite{Salton:2016qpp}.]. Consider the n-partite division of the boundary $\partial M$, with regions $A_{i},B_{j},C_{k},...$, where the boundaries are labeled by representations $i,j,k,...$ of $SU(N)_{1}$. Recall from sec. \ref{sec:3}, that the fusion tensor for $SU(N)_{1}$ is
\begin{equation}
 N_{ijk}, \quad \text{with } i+j+k=0 \mod N
 \end{equation}

 \subsection{Bipartite entropy}
Represent the bipartition of $\partial M$, with boundary regions $A_{i}$ and $B_{j}$ $i,j=0$ to $N-1$, which are representations $(i,j)$ of $SU(N)_{1}$ describing qudit Bell states. The flat entanglement spectrum implies that the entropies are independent of the Renyi parameter. In analogy with equation 9 and Figure 4 of Salton, et. al. \cite{Salton:2016qpp}.
 
 \begin{equation}
 S(A) = -\log\frac{Z(-2M\cup_{f}2M)}{Z(-M\cup_{\partial M}M)^2}
 \end{equation}
 where
 \begin{equation}\label{eq:Z}
 Z(-2M\cup_{f}2M) = N^3
 \end{equation}
 and
 \begin{equation}
 Z(-M\cup_{\partial M}M) = N^2
 \end{equation}
so that
\begin{equation}
 S(A) =\log N
 \end{equation}
 Since $A\cup B$ is a pure state
 \begin{equation}\label{eq:pure}
 S(A)=S(B)
 \end{equation}
 and
 \begin{equation}
 S(AB) =0
 \end{equation}
 
 \subsection{Tripartite entropy}
 Represent $\partial M$ by a tripartite divisor of $\partial M$, divided by regions denoted by 1) $A_{j},B_{j},C_{k}$. The density matrix for region $A\cup B$ is
 \begin{equation}\label{eq:bi}
 \rho_{AB} = \frac{1}{N}\sum_{j=0}^{N-1}\{\ket{jj}\bra{jj}\}_{AB}
 \end{equation}
 which describes qudit Bell-pairs. The tripartite entanglement for the regions 2) $A_{j},B_{j},C_{j}$ is described by the qudit state $\ket{GHZ_{3}}$ with the density matrix
  \begin{equation}\label{eq:tri}
 \rho_{ABC} = \frac{1}{N}\sum_{j=0}^{N-1}\{\ket{jjj}\bra{jjj}\},
 \end{equation}
 which is a stabilizer state.
 
 Any pure tripartite state for a tensor network for large $N$ $SU(N)_{1}$ is equivalent to bipartite maximally entangled state [c.f. \eqref{eq:bi}] and a tripartite GHZ$_{3}$ state [c.f. \eqref{eq:tri}]. There are no $W_{3}$ states! This is analogous to the result of Nezami and Walter \cite{Nezami:2016zni}. Since $ABC$ is a pure state,
 \begin{equation}
 S(ABC)=0
 \end{equation}
 \begin{equation}
 S(AB) = S(C)
 \end{equation}
 From \eqref{eq:Z}, the bipartite entropies appropriate to \eqref{eq:pure} are
 \begin{equation}
 S(A)=S(B)=S(C)=\log N.
 \end{equation}
 
 The tripartite entropies at large $N$ are computed in analogy with Nezami and Walter \cite{Nezami:2016zni}, and satisfies
 \begin{equation}
 g = \frac{1}{\log N} \{S(A)+S(B)+S(C)\}+\log N \frac{Z(-3M\cup_{f}3M)}{Z(-M\cup_{\partial M} M)^3}
 \end{equation}
 which gives $g=1$ as the number of $N$-dimensional GHZ$_{3}$ states.
 
 The mutual information at large $N$ is
 \begin{equation}
 \begin{split}
 I_{2}(A:B) &= S(A)+S(B)-S(AB)\\
 &=S(A)+S(B)-S(C)\\
 &=\log N +\mathcal{O}(1)
 \end{split}
 \end{equation}
 From Nezami and Walter \cite{Nezami:2016zni}
 \begin{equation}
 I_{2}(A:B) = 2c+g
 \end{equation}
 which implies that
 \begin{equation}
 2c = \log N
 \end{equation}
 at large $N$ where $c$ is the number of maximally entangled pairs. Further
 \begin{equation}
 I_{2}(A:B) + I_{2}(B:C) + I_{2}(C:A) = 3\log N
 \end{equation}
 at large $N$, and
 \begin{equation}
 I_{2}(A:BC) = S(A)+S(B)-S(ABC) =2\log N
 \end{equation}
 for a pure ($ABC$) system. Therefore
 \begin{equation}
 I_{2}(A:BC)-I_{2}(A:B) = \log N
 \end{equation}
 The tripartite information
 \begin{equation}
 I_{3} = I_{2}(A_{1}:A_{2})+I_{2}(A_{1}:A_{3})-I_{2}(A_{1}:A_{2}A_{3}) = 0
 \end{equation}
 at large $N$, so that MMI$=0$ at large $N$.
 
 \section{Concluding Remarks}\label{sec:5}
There are several issues related to this paper which deserve further investigation. It would be instructive to study the quantum entropy cones of Chern-Simons $SU(N)_{1}$ and $SU(N)_{K}$; $K\geq2$ in more detail. For the latter we argued that the topological entropy cone is properly contained in the stabilizer entropy cone. In that connection, it would be useful to have an analogue of Theorem 1 of Salton, et. al. \cite{Salton:2016qpp} for $SU(N)_{K}$. That is, for which values of $N$ and $K$ can one only prepare stabilizer states, and for which cases can one prepare any stabilizer state in the n-torus Hilbert space?
 
 It is known that tensor networks only produce stabilizer states \cite{Nezami:2016zni}, which have flat Renyi entropies at large $N$, as was discussed in section \ref{sec:4}. One should note that flat Renyi entropy is not a general feature of Chern-Simons or WZW theories. As an example the Renyi entropy for $SU(N)_1$ WZW theory on a branched rectangular torus of spatial size $L$ and Euclidean time $\beta$, in the small interval limit, is \cite{Schnitzer:2015ira}
 \begin{equation}
 S_{n}=\frac{c}{12}(1+\frac{1}{n})\log(\frac{L}{l})
 \end{equation}
 where $c$ is the central charge of the WZW model. Noteworthy is that $S_{n}$ is not $n$ independent and it exhibits the singularity originating from the branch-cut \cite{Kitaev:2005dm}. By contrast, in the body of this paper, Chern-Simons theories were considered without boundaries, so that $S_{n}$ did not have the singularities\footnote{This distinction was emphasized to us by M. Headrick}.
 
Stabilizer states alone cannot be used to construct a universal quantum computer. One strategy for qudits is to have a generalization of the phase, Hadamard, and $C_{NOT}$ gates. For $SU(N)_{1}$ these are presented in section \ref{sec:3} and in \cite{Schnitzer:2019icr}. Then one can add a generalization of a Toffli (qudit) gate or a hard T-gate. As an example a qutrit version of the T-gate is \cite{White:2020zoz} 
 \begin{equation}
 T = 
 \begin{pmatrix}
 \xi&0&0\\
 0&1&0\\
 0&0&\xi^{-1}
 \end{pmatrix}
 \end{equation}
 where 
 \begin{equation}
 \xi = e^{\frac{2\pi i}{9}}.
 \end{equation}
 The qudrit generalization of the hard $\frac{\pi}{8}$ T-gate operator is also possible \cite{1999,Howard_2012,Prakash_2018,White:2020zoz}.
 
 A different strategy is presented for a $SU(2)_{3}$ universal quantum computer \cite{freedman2000modular}, and its $SU(3)_{2}$ level-rank dual \cite{Schnitzer:2018aef}. It would be interesting to generalize these constructions.
 
\acknowledgments
We thank M. Headrick and B. Swingle for their comments. We are grateful to Isaac Cohen and Jonathan Harper for their aid in preparing the manuscript.

\bibliographystyle{JHEP}
\bibliography{Howard_Chern}
\end{document}